%% file: na59-sos.tex
\documentclass[aps,pra,twocolumn,showpacs,showkeys,superscriptaddress]{revtex4}

\usepackage{graphicx}
\usepackage{dcolumn}
\usepackage{bm}

\begin{document}


\title{Measurement of Coherent Emission and Linear Polarization of Photons 
   by Electrons in the Strong Fields of Aligned Crystals}



\input{na59-sos-authors.tex}

\noaffiliation

\date{\today}

\begin{abstract}
We present new results regarding the features of high energy photon
emission by an electron beam of 178\,GeV penetrating a 1.5\,cm thick
single Si crystal aligned at the Strings-Of-Strings\,(SOS) orientation.
This concerns a special case of coherent bremsstrahlung where the electron
interacts with the strong fields of successive atomic strings in a plane
and for which the largest enhancement of the highest energy photons is
expected. The polarization of the resulting photon beam was measured by
the asymmetry of e$^+$e$^-$ pair production in an aligned diamond crystal
analyzer. By the selection of a single pair the energy and the
polarization of individual photons could be measured in an the environment
of multiple photons produced in the radiator crystal. Photons in the high
energy region show less than 20\% linear polarization at the 90\%
confidence level.
\end{abstract}

\pacs{32.80, 34.80, 61.85.+p, 03.65.Sq, 78.70.-g, 13.88.+e, 41.75.Ht}
\keywords{Single Crystal, Coherent Bremsstrahlung, Channeling 
Radiation, Polarised Photons}

\maketitle


\section{\label{sec:intro}Introduction}

Interest in the generation of intense, highly polarized high energy photon
beams~\cite{proposal,e159} comes in part from the need to investigate the
polarized photo-production mechanisms. For example, the so-called ``spin
crisis of the nucleon'' and its connection to the gluon polarization has
attracted much attention~\cite{compass}. Future experiments will require
intense photon beams with a high degree of polarization.  The radiation
emitted by electrons passing through oriented single crystals is important
for these purposes. The coherent bremsstrahlung\,(CB) of high energy
unpolarized electrons is a well established and widely applied technique
for producing intense photon beams with a high degree of linear
polarization. The coherence arises in this case due to crystal effects
which become pronounced when the electron incidence angle with respect to
a major plane is small. The resulting CB radiation differs from incoherent
bremsstrahlung\,(ICB) in an amorphous target in that the cross section is
substantially enhanced and relatively sharp coherent peaks appear in the
photon spectrum. The position of these peaks can be tuned by adjusting the
electron beam incidence angle with respect to the major plane of the
lattice.

There is another less well known method of producing greater enhancement
as well as a harder photon spectrum than the CB case. This is achieved by
selecting a very specific electron incident angle with respect to the
crystal. If the electron beam is incident very close to the plane (within
the planar channelling critical angle) and also closely well aligned to a
major axis (but beyond the axial channelling critical angle), then the
electron interacts dominantly with successive atomic strings in the plane.
This orientation had been aptly described by the term
``String-Of-Strings"\,(SOS) by Lindhard, a pioneer of beam-crystal
phenomena~\cite{lindhard}. The NA43 Collaboration has studied the
radiation emitted by electrons incident in the SOS orientation. The
reference~\cite{kirsebom01} and references therein are an account of this
study, as well as many other related effects. There remained the issue of
the polarisation of the SOS radiation. Polarisation measurements have been
reported~\cite{kirsebom99} which could be consistent with substantial
polarisation of the hard component of SOS radiation. However the ability
to distinguish clearly a single photon spectrum from the total radiated
energy spectrum was not yet developed for that measurement. In this paper,
a new study of SOS-produced high energy photon beams is reported in which
we were able to study the beam on a photon-by-photon basis, and measure
both the enhancement and the linear polarisation as a function of photon
energy.

\section{\label{theory}Theoretical Description}

The CB mechanism produces linearly polarized photons in a selected energy
region when the crystal type, its orientation with respect to the electron
beam, and the electron energy are appropriately chosen. In the so-called
point effect\,(PE) orientation of the crystal the direction of the
electron beam has a small angle with respect to a chosen crystallographic
plane and a relatively large angle with the crystallographic axes that are
in that plane. For this PE orientation of the single crystal only one
reciprocal lattice vector contributes to the CB cross section. The CB
radiation from a crystal aligned in this configuration is more intense
than the ICB radiation in amorphous media and a high degree of linear
polarization can be achieved~\cite{termisha}. The PE orientation of the
crystal was used in a previous NA59 experiment, where a large linear
polarization of high energy photons was measured. The photons had been
produced by an unpolarized electron beam. The conversion of the linear
polarisation to circular polarization induced by a birefringent effect in
an aligned single crystal was also studied~\cite{na59-1,na59-2}.

The character of the radiation, including its linear polarization, is
changed when the direction of the electron (i) has a small angle with a
crystallographic axis and (ii) is parallel with the plane that is formed
by the atomic strings along the chosen axes.  This is the so-called SOS
orientation. It produces a harder photon spectrum than the CB case because
the coherent radiation arises from successive scattering off the axial
potential, which is deeper than the planar potential. The radiation
phenomena in single crystals aligned in SOS mode have been under active
theoretical investigation since the NA43 collaboration discovered, for the
first time, two distinct photon peaks, one in the low energy region and
one in the high energy region of the radiated energy spectrum for about
150\,GeV electrons traversing a diamond crystal~\cite{new-effect}. It was
established that the hard photon peak was a single photon peak. However,
the radiated photons were generally emitted with significant multiplicity
in such a way that a hard photon would be accompanied by a few low energy
photons. It will be seen later that two different mechanisms are
responsible for the soft and the hard photons. In the former case, it is
planar channelling\,(PC) radiation, while in the latter case, it is SOS
radiation.

The issue of the polarisation of SOS radiation also came into question.
Early experiments with electron beams of up to 10\,GeV in single crystals
showed a smaller linear polarization of the more intense radiation in the
SOS orientation than in the PE orientation (see~\cite{saenz} and
references therein). The first measurements of linear polarization for
high energy photons ($E_{\gamma} \approx 50-150$\,GeV)  were consistent
with a high degree of linear polarization of the radiated
photons~\cite{kirsebom99}. At this stage the theoretical prediction of the
SOS hard photon polarisation was unresolved. However, it was clear that
the photons emitted by the PC mechanism would be linearly polarised. This
experiment therefore could not be considered conclusive, as the
polarimeter recorded the integral polarisation for a given radiated
energy, which was likely to have a multi-photon character. The extent to
which pile-up from the low energy photons perturbed the high energy part
of the total radiated energy spectrum was not resolved. These results
therefore required more theoretical and experimental investigation.

A theory of photon emission by electrons along the SOS orientation of
single crystals has since been developed. The theory takes into account
the change of the effective electron mass in the fields due to the
crystallographic planes and the crossing of the atomic strings~\cite{bks}.
The authors show that the SOS specific potential affects the high energy
photon emission and also gives an additional contribution in the low
energy region of the spectrum.  In Refs.~\cite{simon,strakh} the linear
polarization of the emitted photons was derived and analysed for different
beam energies and crystal orientations. The predicted linear polarization
of hard photons produced using the SOS orientation of the crystal is small
compared to the comparable case using the PE orientation of the crystal.
On the other hand, the additional soft photons produced with SOS
orientation of the crystal are predicted to exhibit a high degree of
polarization.

The emission mechanism of the high energy photons is CB connected to the
periodic structure of the crystal~\cite{termisha}.

The peak energy of the CB photons, $E_\gamma$, is determined from the
condition ( the system of units used here has $\hbar={\rm c}=1$ ),
\begin{equation}
\frac{1}{|q_{\Vert}|} = 2 \lambda_c \gamma \frac{E_0-E_\gamma}{E_\gamma}~,
\end{equation}
where $|q_{\Vert}|$ is the component of the momentum recoil, $\mathbf{q}$,
parallel to the initial electron velocity and the other symbols have their
usual meanings. Recall, in a crystal possible values of $\mathbf{q}$ are
discrete: $\mathbf{q}=\mathbf{g}$~\cite{termisha}, where $\mathbf{g}$ is a
reciprocal lattice vector. The minimal reciprocal lattice vector giving 
rise to the main CB peak in both the PE and the SOS orientations is given 
by
\begin{equation}
|g_{\Vert}|_{min} = \frac{2\pi}{d}\Theta.
\end{equation}

For the PE orientation, $d$ is the interplanar distance and $\Theta=\psi$,
the electron incident angle with respect to the plane. For the SOS
orientation $d$ is the spacing between the axes (strings)  forming the
planes, and $\Theta=\theta$, the electron incident angle with respect to
the axis. The position of the hard photon peak can be selected by 
simultaneous solution of the last two equations,
\begin{equation}
\Theta =\frac{d}{4\pi\gamma\lambda_c}\frac{E_{\gamma}}{E_0-E_{\gamma}}.
\end{equation}

With the appropriate choice of $\theta=\Theta$ the intensity of the SOS
radiation may exceed the ICB radiation by an order of magnitude.

When a thin silicon crystal is used with an electron beam of energy $E_0
=178$\,GeV incident along the SOS orientation, within the $(110)$ plane
and with an angle of $\theta=0.3$ mrad to the $<100>$ axis, the hard
photon peak position is expected at $E_{\gamma}=129$\,GeV.

In the current experiment, a 1.5\,cm thick silicon crystal was used in the
SOS orientation with the electron beam ($E_0 =178$\,GeV) incident within
the $(110)$ plane with an angle of $\theta=0.3$ mrad to the $<100>$ axis.
This gives the hard photon peak position at $x_{max}=0.725$. This
corresponds to the photon energy $E_{\gamma}=129$\,GeV. Under this
condition the radiation is expected to be enhanced by about a factor 30
with respect to the ICB for a randomly oriented crystalline Si target.

The coherence length determines the effective longitudinal dimension of
the interaction region for the phase coherence of the radiation process:
\begin{equation}
l_{coh} = \frac{1}{|q_{\Vert}|}.
\end{equation}

The radiation spectrum with the crystal aligned in SOS orientation has in
addition to the CB radiation a strong component at a low energy which is
characteristic of PC radiation. As the electron direction lines up with a
crystallographic plane in the SOS orientation, the planar channelling
condition is fulfilled. For channelling radiation the coherence length is
much longer than the interatomic distances and the long range motion,
characteristic of planar channelled electrons, becomes dominant over short
range variations with the emission of low energy photons. Theoretical
calculations~\cite{strakh,armen} predict a more intense soft photon
contribution with a high degree of linear polarization of up to 70\%.

\begin{figure}[ht]
\includegraphics[scale=0.433]{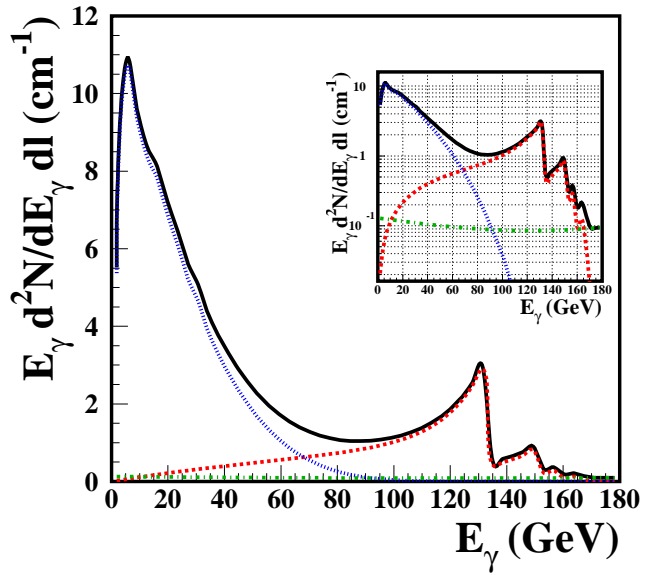}
\caption{\label{F:Strak-1b} Photon power yield per unit of thickness, 
    $E_\gamma d^2N/dE_\gamma dl$, for a thin silicon crystal in the SOS 
    orientation for a $E_0 =178$\,GeV electron beam incident within the 
    $(110)$ plane and at an angle of $\theta=0.3$\,mrad to the $<100>$ 
    axis. At low energy the PC radiation dominates and at high energies 
    the SOS radiation peaks. The solid curve represents the total of the 
    contributions from (green dash-dotted)\,ICB, (blue dotted)\,PC, and 
    (red dashed)\, SOS radiation. vThe insert is a logarithmic 
    representation and shows the flat incoherent contribution and the 
    enhancement with a factor of about 30 for SOS radiation at 129\,GeV.}
\end{figure}

The simulation of the enhancements of both the low energy and the high
energy components of the radiation emission for the SOS orientation under
conditions applicable to this experiment are presented in
Fig.~\ref{F:Strak-1b}.

\section{\label{setup}Experimental Setup}

The NA59 experiment was performed in the North Area of the CERN SPS, where
unpolarized electron beams with energies above 100\,GeV are available. We
used a beam of 178\,GeV electrons with angular divergence of
$\sigma_{x'}=48\,\mu$rad and $\sigma_{y'}=35\,\mu$rad in the horizontal
and vertical plane, respectively.

\begin{figure*}[ht]
\includegraphics[scale=0.624]{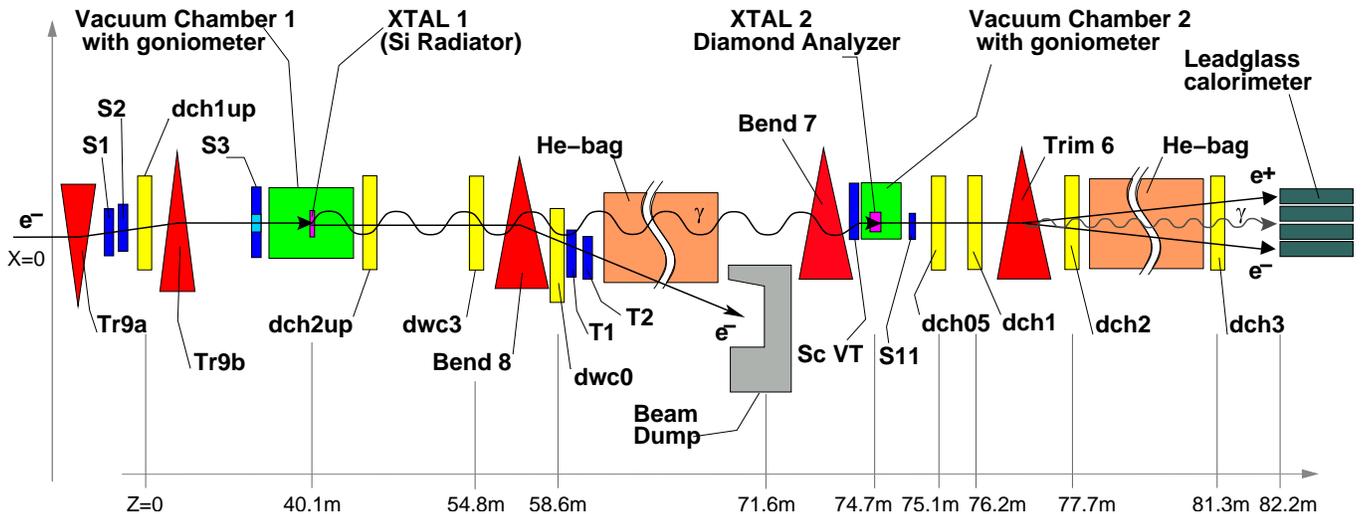}
\caption{\label{F:setup} NA59 experimental setup.}
\end{figure*}

The experimental setup shown in Fig.~\ref{F:setup} was also used to
investigate the linear polarization of CB and birefringence in aligned
single crystals~\cite{na59-1,na59-2}. This setup is ideally suited for
detailed studies of the photon radiation and pair production processes in
aligned crystals.

The main components of the experimental setup are: two goniometers with
crystals mounted inside vacuum chambers, a pair spectrometer, an electron
tagging system, a segmented leadglass calorimeter, wire chambers, and
plastic scintillators. In more detail a 1.5\,cm thick Si crystal can be
rotated in the first goniometer with 2\,$\mu$rad precision and serves as
radiator. A multi-tile synthetic diamond crystal on the first goniometer
can be rotated with 20\,$\mu$rad precision and is used as the analyzer of
the linear polarization of the photon beam.

The photon tagging system consists of a dipole magnet B8, wire chamber
dwc0, and scintillators T1 and T2. Given the geometrical acceptances and
the magnetic field, the system, tags the radiated energy between 10\% and
90\% of the electron beam energy. Drift chambers dch1up, dch2up, and delay
wire chamber dwc3 define the incident and the exit angle of the electron
at the radiator.

The e$^+$e$^-$ pair spectrometer consists of dipole magnet Trim 6 and of
drift chambers dch05, dch1, dch2, and dch3. The drift chambers measure the
horizontal and vertical positions of the passing charged particles with
100\,$\mu$m precision. Together with the magnetic field in the dipole this
gives a momentum resolution of $\sigma_p/p^2=0.0012$ with $p$ in units
of\,GeV/c. The pair spectrometer enables the measurement of the energy of
a high energy photon, $E_\gamma$, in a multi-photon environment. Signals
from the plastic scintillators S1, S2, S3, T1, T2, S11 and veto detector
ScVT provide several dedicated triggers.

The total radiated energy $E_{tot}$ is measured in a 12-segment array of
leadglass calorimeter with a thickness of 24.6 radiation lengths and a
resolution of $\sigma_E=0.115~\sqrt{E}$ with $E$ in units of\,GeV.  A
central element of this leadglass array is used to map and to align the
crystals with the electron beam.

A detailed description of the NA59 experimental apparatus can be found in
reference~\cite{na59-1}.

\section{Results and Discussion}

The experiment can be divided in two parts: (A) production of the photon
beam by the photon radiation of the 178\,GeV electron beam in the Si
radiator oriented in the SOS mode and (B) measurement of the linear
polarization by using diamond crystals as analyzers. Prior to the
experiment Monte Carlo\,(MC) simulations were used to estimate the photon
yield, the radiated energy, and the linear polarization of the photon beam
and we optimized the orientation of the crystal radiator.  The MC
calculations also included the crystal analyzer to estimate the asymmetry
of the e$^+$e$^-$ pair production. The simulations further included the
angular divergence of the electron beam, the spread of 1\% in the beam
energy, and the generation of the electromagnetic shower that develops in
oriented crystals. To optimize the processing time of the MC simulation,
energy cuts of 5\,GeV for electrons and of 500\,MeV for photons were
applied.

\subsection{Photon Beam}

We used a beam angle of $\theta=0.3$\,mrad to the $\langle 100 \rangle$
axis in the $(110)$ plane of the 1.5\,cm thick Si crystal which is the
optimal angle for a high energy SOS photon peak at 129\,GeV (see
Fig.~\ref{F:Strak-1b}). As is mentioned above, the radiation probability
with a thin radiator is expected to be 30 times larger at that energy than
the Bethe-Heitler\,(ICB) prediction for randomly oriented crystalline Si.

\begin{figure}[ht]
\includegraphics[scale=0.433]{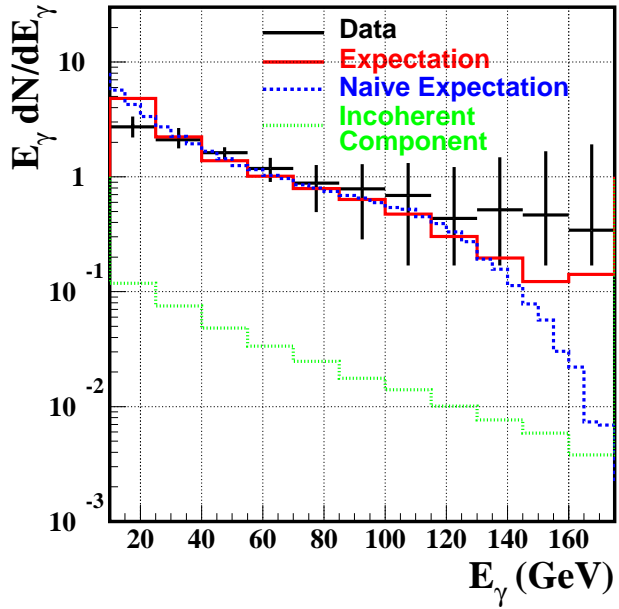}
\caption{\label{F:sps} Photon power yield, $E_\gamma dN/dE_\gamma$, as a 
    function of the energy $E_\gamma$ of individual photons radiated by an 
    electron beam of 178\,GeV in the SOS-aligned 1.5\,cm Si crystal. The 
    black crosses are the measurements with the pair spectrometer, the 
    vertical lines represent the errors including the uncertainty in the 
    acceptance of the spectrometer. The (red solid) histogram represent 
    the MC prediction for our experimental conditions. The (green dotted) 
    represent the small contribution due to incoherent interactions. For 
    completeness, we also show the theoretical predictions if the 
    experimental effects are ignored  (blue dashed).}
\end{figure}

However, there are several consequences for the photon spectrum due to the
use of a 1.5 cm thick crystal For the chosen orientation of the Si
crystal, the emission of mainly low energy photons from planar coherent
bremsstrahlung (PC) results in a total average photon multiplicity above
15. And the most probable radiative energy loss of the 178 GeV electrons
is expected to be 80\%. The beam energy decreases significantly as the
electrons traverse the crystal. The peak energy of both SOS and PC
radiation also decreases with the decrease in electron energy.
Consequently, the SOS radiation spectrum is not peaked at the energy for a
thin radiator, but becomes a smooth energy distribution. Clearly, many
electrons may pass through the crystal without emitting SOS radiation and
still lose a large fraction of their energy due to PC and ICB. Hard
photons emitted in the first part of the crystal that convert in the later
part do not contribute anymore to the high energy part of the photon
spectrum.  A full Monte Carlo calculation is necessary to propagate the
predicted photon yield with a thin crystal, as shown in
Fig.~\ref{F:Strak-1b} for 178 GeV electrons, to the current case with a
1.5\,cm thick crystal.

This has been implemented for the measured photon spectrum shown in
Fig.~\ref{F:sps}.  We see that the measured SOS photon spectrum shows a
smoothly decreasing distribution.  The low energy region of the photon
spectrum is especially saturated, due to the abundant production of low
energy photons. Above 25\,GeV however, there is satisfactory agreement
with the theoretical Monte Carlo prediction, which includes the effects
mentioned above.

\begin{figure}[ht]  
\includegraphics[scale=0.433]{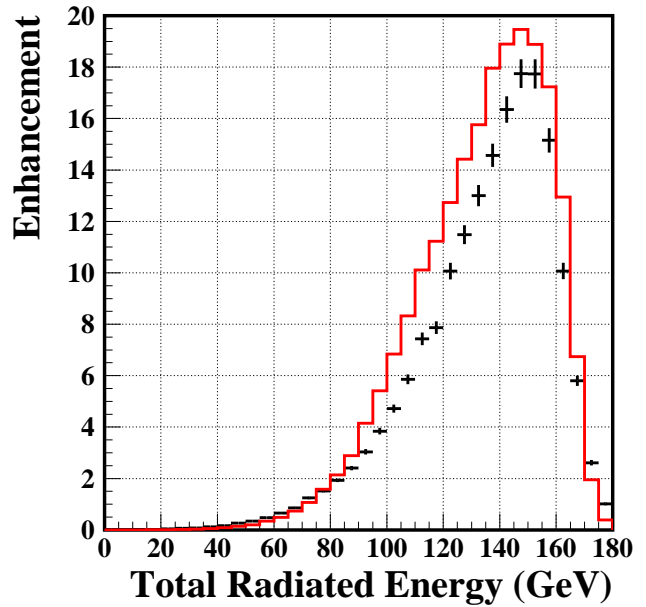}
\caption{\label{F:enh} Enhancement of the intensity with respect to the 
     Bethe-Heitler\,(ICB) prediction for randomly oriented polycrystalline 
     Si as a function of the total radiated energy $E_{tot}$ in the 
     SOS-aligned Si crystal by 178\,GeV electrons. The black crosses are 
     the measurements and the red histogram represent the MC prediction.}
\end{figure}

The enhancement of the emission probability compared to the ICB prediction
is given in Fig.~\ref{F:enh} as a function of the total radiated energy as
measured in the calorimeter. The maximal enhancement is about a factor of
18 at 150\,GeV and corresponds well with the predicted maximum of about 20
at 148\,GeV. This is a multi-photon spectrum measured with the photon
calorimeter. The peak of radiated energy is situated at 150\,GeV, which
means that each electron lost about 80\% of its initial energy due to the
large thickness of the radiator. This means that the effective radiation
length of the oriented single crystal is several times shorter in
comparison with the amorphous target. The low energy region is depleted
due to the pile-up of several photons.

\indent

The expected linear polarization is shown in Fig.~\ref{F:SOS-pol} as a
function of photon energy. It is well known that channelling radiation in
single crystals is linearly polarized~\cite{Adishchev,Vorobyov} and the
low energy photons up to 70\,GeV are also predicted to be linearly
polarized in the MC simulations. High energy photons are predicted with an
insignificant polarization.

\subsection{Asymmetry Measurement}

In this work, the photon polarization is always expressed using the
Stoke's parametrization with the Landau convention, where the total
elliptical polarization is decomposed into two independent linear
components and a circular component. In mathematical terms, one writes:

\begin{widetext}
\begin{equation}
P_{\hbox {linear}}=\sqrt{\eta _{1}^{2}+\eta _{3}^{2}},
\quad \; P_{\hbox {circular}}=\sqrt{\eta _{2}^{2}},
\quad \; P_{\hbox {total}}=\sqrt{P_{\hbox {linear}}^{2}+P_{\hbox
{circular}}^{2}} \quad .
\label{eq:pol-def}
\end{equation}
\end{widetext}

\begin{figure}[ht]
\includegraphics[scale=0.433]{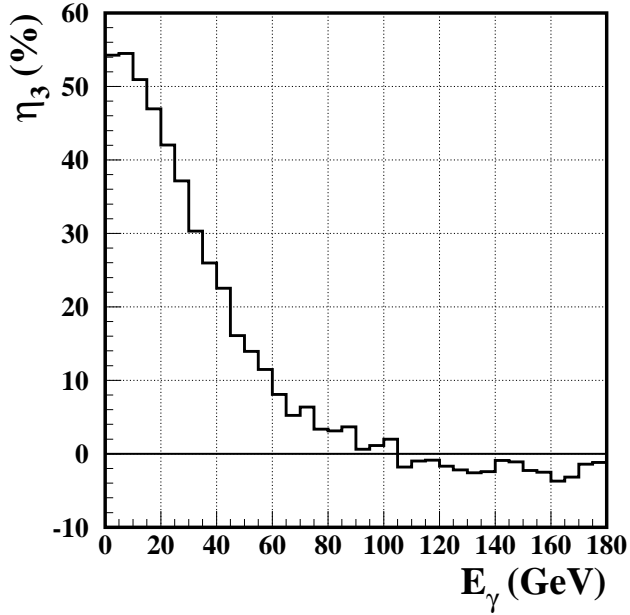}
\caption{\label{F:SOS-pol} Expected linear polarization as a function of 
    the energy $E_\gamma$ of the photons produced in the SOS-aligned Si 
    crystal by 178\,GeV electrons.}
\end{figure}

The radiator angular settings were chosen to have the total linear
polarization from the SOS radiation purely along $\eta _{3}$, that is
$\eta _{1}=0$. The $\eta _{2}$ component is also zero because the electron
beam is unpolarized. The expected $\eta _{3}$ component of the
polarization shown is in Fig.~\ref{F:SOS-pol}.

In order to determine the linear polarization of the photon beam the
method proposed in reference~\cite{barbiellini} with an oriented crystal
was chosen. This method of measurement of the linear polarization of high
energy photons is based on coherent e$^+$e$^-$ pair production\,(CPP) in
single crystals which depends on the orientation of the reciprocal lattice
vector and the linear polarization vector. Thus, the dependence of the CPP
cross section on the linear polarization of the photon beam makes an
oriented single crystal suitable as an efficient polarimeter for high
energy photons.

The basic characteristic of the polarimeter is the analyzing power $R$ of
the analyzer crystal~\cite{barbiellini}. By choosing the appropriate
crystal type and its orientation a maximal analyzing power can be
obtained. The relevant experimental quantity is the asymmetry $A$ of the
cross sections $\sigma (\gamma \rightarrow e^+e^-)$ for parallel and
perpendicular polarization, where the polarization direction is defined
with respect to a particular crystallographic plane of the {\em analyzer}
crystal. This asymmetry is related to the linear polarization of the
photon beam, $P_{\rm linear}$, through:
\begin{equation}
A  \equiv   \frac{\sigma (\gamma _{\perp }\rightarrow e^{+}e^{-})-\sigma
(\gamma _{\parallel }\rightarrow e^{+}e^{-})}{\sigma (\gamma _{\perp }  
\rightarrow e^{+}e^{-})+\sigma (\gamma _{\parallel }\rightarrow
e^{+}e^{-})}
=R \times P_{\rm linear}.
\label{eq:asym}
\end{equation}
The analyzing power $R$ corresponds to the asymmetry expected for photons
that are 100\% linearly polarized perpendicular to the chosen
crystallographic plane.

Denoting the number of e$^+$e$^-$ pairs produced in perpendicular and
parallel cases by $p_{1}$ and $p_{2}$, and the number of the normalisation
events in each case by $n_{1}$ and $n_{2}$, respectively, the measured
asymmetry can be written as:
\begin{equation}
A=\frac{p_{\perp }/n{\perp } - p_{\parallel }/n_{\parallel }}{p_{\perp 
}/n_{\perp } + p_{\parallel }/n_{\parallel }},
\label{eq:asy-meas}
\end{equation}
where $p$ and $n$ are acquired simultaneously and therefore correlated.  
Further details of this method, as well as refinements to enhance the
analyzing power $R$ by using kinematic cuts on the pair spectra, may be
found in reference~\cite{na59-1}.

The existence of a strong anisotropy for the channelling of the e$^+$e$^-$
pairs during their formation is the reason for the polarization dependent
CPP cross section of photons passing through oriented crystals. This means
that perfect alignment along a crystallographic axis is not an efficient
analyzer orientation due to the approximate cylindrical symmetry of the
crystal around atomic strings. However, for small angles of the photon
beam with respect to the crystallographic symmetry directions the
conditions for the formation of the e$^+$e$^-$ pairs prove to be very
anisotropic. As it turns out, the orientations with the highest analyzing
power are those where the e$^+$e$^-$ pair formation zone is not only
highly anisotropic but also inhomogeneous with maximal fluctuations of the
crystal potential along the electron path. At the crystallographic axes
the potential is largest and so are the fluctuations. These conditions are
related to the ones of the SOS orientation: (i) a small angle to a
crystallographic axis to enhance the pair production (PP) process by the
large fluctuations and (ii) a smaller angle to the crystallographic plane
to have a long but still anisotropic formation zone for CPP.

\begin{figure}[ht]
\includegraphics[scale=0.433]{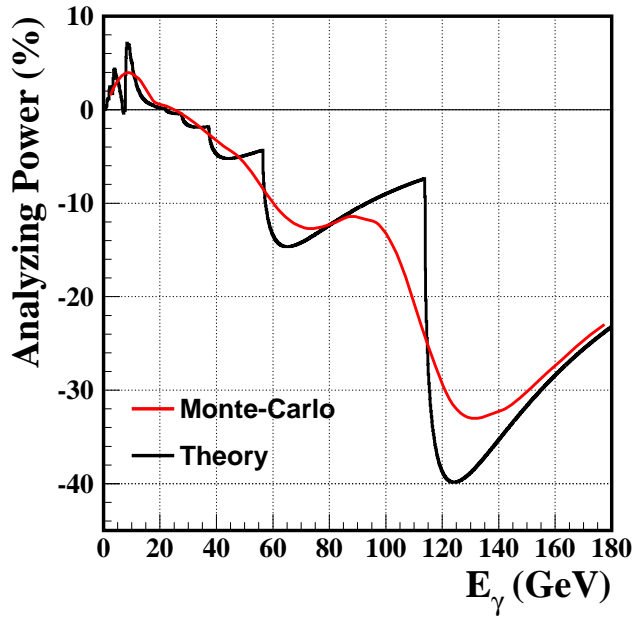}
\caption{\label{F:anpow} Analyzing power $R$ with the aligned diamond 
    crystal as a function of the photon energy $E_\gamma$ (black curve) 
    for an ideal photon beam without angular divergence and (red curve) 
    for the Monte Carlo simulation of photons with the beam conditions in 
    the NA59 experiment.}
\end{figure}

In the NA59 experiment we used a multi-tile synthetic diamond crystal as
an analyzer oriented with the photon beam at 6.2 mrad to the
\mbox{$\langle 100 \rangle$} axis and at 465\,$\mu$rad from the $(110)$
plane. This configuration is predicted to have a maximal analyzing power
for a photon energy of 125\,GeV as is shown in Fig.~\ref{F:anpow}. The
predicted analyzing power in the high energy peak region is about 30\%.

\begin{figure}[h]
\includegraphics[scale=0.433]{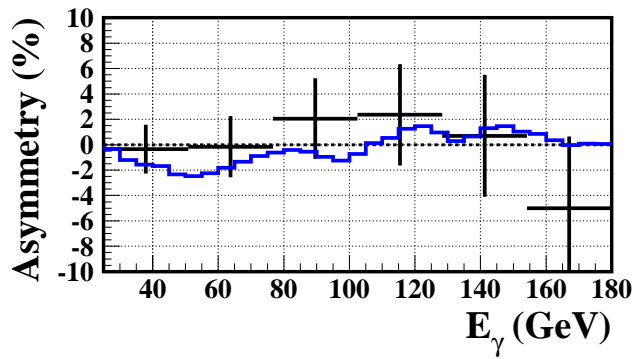}
\caption{\label{F:asy} Asymmetry of the e$^+$e$^-$ pair production in the 
     aligned diamond crystal as a function of the photon energy $E_\gamma$ 
     which is measured to determine the $P_1$ component of the photon 
     polarization in the SOS-aligned Si crystal by 178\,GeV electrons. The 
     black crosses are the measurements and the red histogram represent 
     the MC prediction.}
\end{figure}

The measured asymmetry and the predicted asymmetry are shown in
Fig.~\ref{F:asy}. One can see that the measured asymmetry is consistent
with zero over the whole photon energy range. For the photon energy range
of 100-155\,GeV we find less than 5\% polarization at 0.9 confidence
level. The null result is expected to be reliable as the correct operation
of the polarimeter had been confirmed in the same beam-time in
measurements of the polarisation of CB radiation~\cite{na59-1}. Note, that
the expected asymmetry is small, especially in the high energy range of
120-140\,GeV, where the analyzing power is large, see
Fig.~(\ref{F:anpow}). This corresponds to the expected small linear
polarization in the high energy range, see Fig.~(\ref{F:SOS-pol}).

In contrast to the result of a previous experiment~\cite{kirsebom99}, our
results are consistent with calculations that predict negligible
polarization in the high energy photon peak for the SOS orientation. The
analyzing power of the diamond analyzer crystal in the previous
experiment's~\cite{kirsebom99} setup peaked in the photon energy range of
20-40\,GeV where a high degree of linear polarization is expected. But in
the high energy photon region we expect a small analyzing power of about
2-3\%, also following recent calculation~\cite{simon,strakh}. The constant
asymmetry measured in a previous experiment \cite{kirsebom99} over the
whole range of total radiated energy may therefore not be due to the
contribution of the high energy photons.

From Fig.~\ref{F:SOS-pol} one can expect a large linear polarization for
photons in the low energy range of 20-50\,GeV. However, the analyzing
power was optimized for an photon energy of 125\,GeV and is small in the
region where we expect a large polarization. A different choice of
orientation of the analyzer crystal can move the analyzing power peak to
the low energy range and may be used to measure the linear polarization in
the low energy range.

\section{Conclusion}

We have performed an investigation of both enhancement and polarisation of
photons emitted in the so called SOS radiation. This is a special case of
coherent bremsstrahlung for multi-hundred GeV electrons incident on
oriented crystalline targets, which provides some advantages comparing
with other types of CB orientations. The experimental set-up had the
capacity to deal with the relatively high photon multiplicity and single
photon spectra were measured. This is very important in view of the fact
that there are several production mechanisms for the multiphotons, which
have different radiation characteristics.

We have confirmed the single photon nature of the hard photon peak
produced in SOS radiation.

The issue of the polarisation of the SOS photons had previously not been
conclusively settled. Earlier results in a previous
experiment~\cite{kirsebom99} had indicated that a large polarization might
be obtained for the high energy SOS photons. Our experimental results show
that the high energy photons emitted by electrons passing through the Si
crystal radiator oriented in the SOS mode have a linear polarization
smaller than 20\% at a confidence level of 90\%.

Since the previous experiments, the theoretical situation for the
polarisation of hard SOS photons has also become clearer. Our results
therefore also confirm recent calculations which predict that the linear
polarization of high energy photons created in SOS orientation of the
crystal is small compared to the polarization obtained with the PE
orientation.

Photon emission by electrons traversing single crystals oriented in the
SOS orientation has interesting peculiarities since three different
radiation processes are involved: (1) incoherent bremsstrahlung, (2)
channelling radiation, and (3) coherent bremsstrahlung induced the
periodic structure of the atomic strings in the crystal that are crossed
by the electron. The calculations presented here have taken these three
processes into account, and predict around a 5\% polarization for the high
energy SOS photons. This prediction is consistent with our null
polarization asymmetry measurement for the single photons with energies
above 100\,GeV.

\begin{acknowledgments}

We dedicate this work to the memory of Friedel Sellschop. We express our
gratitude to CNRS, Grenoble for the crystal alignment and Messers DeBeers
Corporation for providing the high quality synthetic diamonds.  We are
grateful for the help and support of N. Doble, K. Elsener and H. Wahl. It
is a pleasure to thank the technical staff of the participating
laboratories and universities for their efforts in the construction and
operation of the experiment.

This research was partially supported by the Illinois Consortium for
Accelerator Research, agreement number~228-1001. UIU acknowledges support
from the Danish Natural Science research council, STENO grant no
J1-00-0568.

\end{acknowledgments}

\bibliography{na59-sos}

\end{document}

%% file: na59-sos-authors.tex
\author{A.~Apyan}
\altaffiliation[Now at: ]{Northwestern University, Evanston, USA}
\affiliation{Yerevan Physics Institute, Yerevan, Armenia}

\author{R.O.~Avakian}
\affiliation{Yerevan Physics Institute, Yerevan, Armenia}

\author{B.~Badelek}
\affiliation{Uppsala University, Uppsala, Sweden}

\author{S.~Ballestrero}
\affiliation{INFN and University of Firenze, Firenze, Italy}

\author{C.~Biino}
\affiliation{INFN and University of Torino, Torino, Italy}
\affiliation{CERN, Geneva, Switzerland}

\author{I.~Birol}
\affiliation{Northwestern University, Evanston, USA}

\author{P.~Cenci}
\affiliation{INFN, Perugia, Italy}

\author{S.H.~Connell}
\affiliation{Schonland Research Centre - University of the Witwatersrand,
Johannesburg, South Africa}

\author{S.~Eichblatt}
\affiliation{Northwestern University, Evanston, USA}

\author{T.~Fonseca}
\affiliation{Northwestern University, Evanston, USA}

\author{A.~Freund}
\affiliation{ESRF, Grenoble, France}

\author{B.~Gorini}
\affiliation{CERN, Geneva, Switzerland}

\author{R.~Groess}
\affiliation{Schonland Research Centre - University of the Witwatersrand,
Johannesburg, South Africa}

\author{K.~Ispirian}
\affiliation{Yerevan Physics Institute, Yerevan, Armenia}

\author{T.J.~Ketel}
\affiliation{NIKHEF, Amsterdam, The Netherlands}

\author{Yu.V.~Kononets}
\affiliation{Kurchatov Institute, Moscow, Russia}

\author{A.~Lopez}
\affiliation{University of Santiago de Compostela, Santiago de Compostela,
Spain}

\author{A.~Mangiarotti}
\affiliation{INFN and University of Firenze, Firenze, Italy}

\author{B.~van~Rens}
\affiliation{NIKHEF, Amsterdam, The Netherlands}

\author{J.P.F.~Sellschop}
\altaffiliation[Deceased]{}
\affiliation{Schonland Research Centre - University of the Witwatersrand,
Johannesburg, South Africa}

\author{M.~Shieh}
\affiliation{Northwestern University, Evanston, USA}

\author{P.~Sona}
\affiliation{INFN and University of Firenze, Firenze, Italy}

\author{V.~Strakhovenko}
\affiliation{Institute of Nuclear Physics, Novosibirsk, Russia}

\author{E.~Uggerh{\o}j}
\thanks{Co-Spokeperson}
\affiliation{Institute for Storage Ring Facilities, University of Aarhus,
Denmark}

\author{U.I.~Uggerh{\o}j}
\affiliation{University of Aarhus, Aarhus, Denmark}

\author{G.~Unel}
\affiliation{Northwestern University, Evanston, USA}

\author{M.~Velasco}
\thanks{Co-Spokeperson}
\altaffiliation[Now at: ]{Northwestern University, Evanston, USA}
\affiliation{CERN, Geneva, Switzerland}

\author{Z.Z.~Vilakazi}
\altaffiliation[Now at: ]{University of Cape Town, Cape Town, South Africa}

\affiliation{Schonland Research Centre - University of the Witwatersrand,
Johannesburg, South Africa}

\author{O.~Wessely}
\affiliation{Uppsala University, Uppsala, Sweden}

\collaboration{The NA59 Collaboration}